\documentclass[aps,prl,twocolumn,showpacs,notitlepage,floatfix,longbibliography,superscriptaddress]{revtex4-1}

\usepackage[breaklinks=true]{hyperref}
\hypersetup{colorlinks=true,linkcolor=blue,citecolor=blue,filecolor=blue,urlcolor=blue,pdfstartview=FitH}
\usepackage{breakcites}
\usepackage{amsmath}
\usepackage{amssymb}
\usepackage{graphicx}
\usepackage[normalem]{ulem}
\usepackage{physics}
\usepackage{xcolor}

\newcommand{\sect}[1]{\textsl{#1} ---}

\newcommand{\cD}{\ensuremath{\mathcal{D}}}
\newcommand{\cF}{\ensuremath{\mathcal{F}}}

\newcommand{\mk}{\mathbf{k}}

\newcommand{\mh}{\hat{\ensuremath{\mathcal{H}}}}
\newcommand{\mhsp}{\hat{\ensuremath{\mathcal{H}}}_\text{sp}}
\newcommand{\mhi}{\hat{\ensuremath{\mathcal{H}}}_\text{int}}

\newcommand{\mhq}{\hat{\ensuremath{\mathcal{H}}}^{\{{\bf q}\}}}

\newcommand{\dtf}{\delta_\text{f}}
\newcommand{\dtd}{\delta_\text{d}}

\newcommand{\heps}{\hat{\ensuremath{\varepsilon}}}

\newcommand{\ha}{\hat{a}}
\newcommand{\hb}{\hat{b}}
\newcommand{\hc}{\hat{c}}
\newcommand{\hd}{\hat{d}}
\newcommand{\hf}{\hat{f}}
\newcommand{\hg}{\hat{g}}

\newcommand{\hn}{\hat{n}}
\newcommand{\hp}{\hat{p}}
\newcommand{\hq}{\hat{q}}

\newcommand{\efb}{E_\text{FB}}

\newcommand{\pksadd}{Max Planck Institute for the Physics of Complex Systems, Dresden D-01187, Germany}
\newcommand{\pcsadd}{Center for Theoretical Physics of Complex Systems, Institute for Basic Science (IBS), Daejeon 34126, Republic of Korea}
\newcommand{\ustadd}{Basic Science Program, Korea University of Science and Technology (UST), Daejeon, Korea, 34113}

\begin{document}

\title{Many-body localization transition from flatband fine-tuning}

\author{Carlo Danieli}
\affiliation{\pksadd}

\author{Alexei Andreanov}
\affiliation{\pcsadd}
\affiliation{\ustadd}

\author{Sergej Flach}
\affiliation{\pcsadd}
\affiliation{\ustadd}

\date{\today}

\begin{abstract}
Translationally invariant flatband Hamiltonians with interactions lead to a many-body localization transition. Our models are obtained from single particle lattices hosting a mix of flat and dispersive bands, and equipped with fine-tuned two--body interactions. Fine-tuning of the interaction results in an extensive set of local conserved charges and a fragmentation of the Hilbert space into irreducible sectors. In each sector, the conserved charges originate from the flatband and act as an effective disorder inducing a transition between ergodic and localized phases upon variation of the interaction strength. Such fine-tuning is possible in arbitrary lattice dimensions and for any many-body statistics. We present computational evidence for this transition with spinless fermions. 
\end{abstract}

\maketitle


\sect{Introduction} 
The celebrated Anderson localization~\cite{anderson1958absence} with additional interactions leads to a novel phase of matter dubbed Many-Body Localization (MBL). Efforts to understand this phase generated an impressive body of work devoted to the study of non equilibrium quantum many-body systems. Following the first pioneering works~\cite{fleishman1980interaction,Altshuler1997quasi,Jacquod1997emergence,Gornyi2005interacting,basko2006metal}, a large number of MBL--related theoretical and experimental studies focused on the interplay of disorder and interaction -- as summarized in~\cite{abanin2017recent,abanin2019colloquium}. Interestingly, a variety of diverse interacting systems were reported to enter MBL phases even in the absence of disorder ~\cite{schiulaz2015dynamics,vanhorssen2015dynamics,pino2016nonergodic,hickey2016signatures,mondaini2017many,schulz2019stark}. This opened an active research quest dedicated to disorder--free MBL. Ergodicity breaking in disorder--free setups can appear due to the splitting of the Hilbert space into exponentially large number of disconnected parts. It is induced by the presence of an extensive number of local conserved quantities. Discussions relate to lattice models endorsing spin-duality relations~\cite{smith2017disorder,smith2017absence,smith2018dynamical} or gauge invariance~\cite{brenes2018many}, a two-dimensional quantum-link network~\cite{karpov2020disorderfree}, and a two-leg compass ladder~\cite{hart2020logarithmic}. Examples of such splitting have been also found in setups without any apparent extensive number of conserved quantities -- {\it e.g.} in systems conserving dipole moments~\cite{sala2020ergdicity} or domain-wall numbers~\cite{Yang2020hilbert}. While first realizations of this phenomenon have been recently emerging ({\it e.g.}~\cite{scherg2020observing}) the above references utilize rather abstract models whose applicability to experimental realizations might be far from trivial.

We use translationally invariant short-range flatband Hamiltonian networks. Geometric frustration in these finetuned systems results in a mix of dispersionless (flat) and dispersive Bloch bands. A hallmark is the existence of compact localized (eigen)states (CLS) spanned over a finite number  $U$ of unit cells. Flatband lattices and CLS have been extensively studied over the last decades~\cite{derzhko2015strongly,leykam2018artificial,leykam2018perspective}, and although the vast majority of results concern single particle problems -- \emph{e.g.} lattice generator schemes~\cite{flach2014detangling,dias2015origami,ramachandran2017chiral,maimaiti2017compact,roentgen2018compact,toikka2018necessary,maimaiti2019universal,maimaiti2021flatband} -- flatbands are progressively entering the realm of quantum many-body physics. Even more importantly, a plethora of experimental studies using an impressive variety of physical platforms were performed, which demonstrate the broad applicability of the finetuning procedure~\cite{leykam2018artificial}. Recently many-particles CLS~\cite{tovmasyan2018preformed,tilleke2020nearest,danieli2020quantum} and flatband-induced quantum scars~\cite{hart2020compact,mcclarty2020disorder,kuno2020flat_qs} have been introduced. Networks which  completely lack single particle dispersion (all bands flat), can completely suppress charge transport with fine-tuned interaction~\cite{danieli2020many,kuno2020flat,orito2020exact}, while adding onsite disorder and interactions leads to conventional MBL features~\cite{Roy2020interplay}. We show that disorder free MBL needs just one flatband and at least one dispersive band when accompanied with a proper interaction finetuning. Our results explain recent reports on MBL-like dynamics for interacting spinless fermions in particular flatband lattices~\cite{daumann2020manybody,khare2020localized}. 


\sect{Setup}
We consider a translationally invariant many-body Hamiltonian 
\begin{gather}
    \label{eq:Ham1}
    \mh = \mhsp + V \mhi\;,\quad \mhsp = \sum_l \hf_l\;,\; \; \mhi = \sum_m \hg_m
\end{gather}
with single particle $\mhsp$ and interaction $\mhi$ parts written as sums of local operators $\hf_l$ and $\hg_m$. The integers $l, m$ label unit cells (with either same or different unit cell choices). Each unit cell contains $\nu$ sites, and the spectrum of $\mhsp$  $\nu$ single particle bands. The local operators $\hf_l,\hg_m$ are given by products of annihilation and creation operators $\hc_{l,a}, \hc_{l,a}^\dagger$ with $1\leq a\leq \nu$.

We consider $\mhsp$ which hosts a flat band $\efb$ while the remaining $\nu-1$ bands are dispersive. Our results generalize to the case of multiple flatbands. Flatbands with short-range hopping have compact localized states (CLS), and we consider the case where these eigenstates form an orthonormal basis. Following Ref.~\onlinecite{flach2014detangling}, the original basis $\hc_{l,a}, \hc_{l,a}^\dagger$ of $\mhsp$ can be recast via local unitary transformations into a new representation $\ha_{l,a},\ha_{l,a}^\dagger$ in which $\mhsp$ turns into a sum of two commuting components $\mhsp =  \mhsp^\text{f} +  \mhsp^\text{d}$. In particular, these components are defined over two disjoint sublattices $\cF$ and $\cD$ formed by the a single particle flatband and the $\nu-1$ dispersive bands respectively. In this \emph{detangled representation}, the flatband component $\mhsp^\text{f}$ in terms of local operators over sublattice $\cF$ reads
\begin{gather}
    \label{eq:Ham_FB}
    \mhsp^\text{f} = \efb \sum_l \ha_{l,1}^\dagger \ha_{l,1} 
    \end{gather} 
The dispersive component $\mhsp^\text{d}$  is expressed in the Bloch basis $\ha_{l,a}^{(\dagger )} = \sum_\mk e^{(-)i \mk l} \hd_{\mk, a}^{(\dagger )}$ for the annihilation (creation) operators $\ha_{l,a}$ ($\ha_{l,a}^\dagger$) for $2\leq a\leq \nu$ in terms of local operators over sublattice $\cD$ 
\begin{gather}
    \label{eq:Ham_DB}
    \mhsp^\text{d}= \sum_{a=2}^{\nu} \sum_\mk E_a(\mk) \ \hd_{\mk, a}^\dagger \hd_{\mk, a} 
\end{gather}
where $\{ E_a(\mk) \}_{a=2}^\nu$ are the dispersive bands of $\mhsp$.

We assume the interaction $\mhi$ in Eq.~\eqref{eq:Ham1} to be two-body -- hence, the local operators $\hg_m$ are written as 
\begin{gather}
    \label{eq:Ham_INT}
    \hg_m = \sum_{\alpha,\beta,\gamma,\delta=1}^\nu J_{\alpha \beta \gamma \delta} \hc_{m,\alpha}^\dagger \hc_{m,\beta} ^\dagger \hc_{m,\gamma} \hc_{m,\delta} + \text{h.c.}
\end{gather}
In the detangled representation of $\mhsp$, the interaction $\mhi$ splits in three components
\begin{gather}
    \label{eq:Ham_INT_split}
    \mhi =  \mhi^\text{f} + \mhi^\text{d} + \mhi^\text{df}
\end{gather}
where (i) the \emph{flatband component} $\mhi^\text{f}$ is defined over sublattice $\cF$ with indices $\alpha, \beta, \gamma, \delta = 1$ in~\eqref{eq:Ham_INT}; (ii) the \emph{dispersive component} $\mhi^\text{d}$ is defined over sublattice $\cD$ with $2\leq \alpha, \beta, \gamma, \delta\leq \nu$ in~\eqref{eq:Ham_INT}; and (iii) the \emph{intra flat--dispersive component} $\mhi^\text{df}$ is defined by all those terms in Eq.~\eqref{eq:Ham_INT} which are not accounted for by either $\mhi^\text{f},\mhi^\text{d}$.

The Hamiltonian $\mhsp^\text{f}$ in Eq.~\eqref{eq:Ham_FB} is formed only by particle number operators $\hn=\ha^\dagger\ha$ and coined Fully Detangled (FD)~\onlinecite{danieli2020many}. Likewise, if we take $J_{\alpha \beta \gamma \delta} = J_{\alpha \beta} \delta_{\alpha,\gamma} \delta_{\beta,\delta}$ in one of the three components $\mhi^\text{f}, \mhi^\text{d}, \mhi^\text{df}$ in Eq.~\eqref{eq:Ham_INT_split} for the correspondent subset of indices, then that component is FD as well -- as a combination of density operators $\hn$ only.

We first consider Hamiltonians $\mh$ in Eq.~\eqref{eq:Ham1} with $\mhi^\text{f}, \mhi^\text{df}$ in Eq.~\eqref{eq:Ham_INT_split} being FD. This condition forbids particles to move within sublattice $\cF$ nor to move from sublattice $\cF$ to $\cD$ and vice versa. Then particles are locked within the flatband component and $\mh$ possesses an extensive set of local conserved quantities $\hq_l = \hn_l = \ha_l^\dagger \ha_l$ for any $l$. Consequently, the relevant Hamiltonian $\mh$ in Eq.~\eqref{eq:Ham1} can be reduced to $\mh = \mhq + \mhsp^\text{f} + V \mhi^\text{f}$. The operators  $\mhsp^\text{f}, \mhi^\text{f}$ depend solely on the conserved quantities $\{\hq_l \}$ and are therefore irrelevant for the particle dynamics. The relevant Hamiltonian
\begin{align}
    \label{eq:Ham_D_eff}
    \mhq &= \mhsp^\text{d} + V \mhi^\text{d} +  \sum_{m, \beta}  \heps_{m,\beta} \hn_{m,\beta}
\end{align}
where $\heps_{m,\beta} = V J_{1 \beta} \hq_{m}$ governs the dynamics of interacting particles in the sublattice $\cD$. The term $\heps_{m,\beta}$ originates from interaction between the flat and dispersive bands -- i.e. the intra flat-dispersive interaction component $\mhi^\text{df}$ -- and it depends on the realization values of conserved quantities $\{\hq_{l}\}$. Hence, the particles locked in the flatband component act as scatterers for the moving particles in the dispersive component, inducing an effective discrete potential $\heps_{m,\beta}$ whose strength is controlled by the interaction strength $V$. 

The Hilbert space of the \emph{full} Hamiltonian $\mh$ in Eq.~\eqref{eq:Ham1} is fragmented: it contains irreducible sectors for any filling fraction $\delta$, which are characterized by the the values of the conserved quantities $\{q_l = \mel{\psi}{\hq_l}{\psi}\}$ -- similarly to \emph{e.g.} Refs.~\cite{smith2017disorder,smith2017absence,smith2018dynamical,brenes2018many,karpov2020disorderfree,hart2020logarithmic}. In a given sector, the wavefunction decomposes into $\ket{\psi} = \ket{\psi_\text{f}}\otimes\ket{\psi_\text{d}}$, where $\ket{\psi_\text{f}} =  \ket{q_1, \dots}$ accounts for $M_\text{f} =  \sum_{l\leq L} q_l$ particles locked in the flatband component with filling fraction $\dtf = M_\text{f}/L$, while $\ket{\psi_\text{d}}$ accounts for the mobile particles with filling fraction $\dtd$ in the dispersive component whose dynamics is governed by $\mhq$~\eqref{eq:Ham_D_eff}. Both filling fractions $\dtf$ and $\dtd$ result in the overall filling fraction $\delta = \frac{1}{\nu}\ \dtf + \frac{\nu-1}{\nu}\ \dtd$. The total number $T_\text{f}$ of sectors depends on both $\dtf$, the system size $L$, and the many-body statistics -- {\it e.g.} for spinless fermions, $T_\text{f} = {{L}\choose{M_\text{f}}}$, while for bosons $T_\text{f} = {M_\text{f}}^{L}$. Indeed, for spinless fermions $q_l =0,1$ while for bosons  $0 \leq q_l \leq M_f$, which consequently yield in these cases different value ranges for the potential $\heps_{m,\beta}$ in Eq.~\eqref{eq:Ham_D_eff} (i.e. different potential strengths).

For a fixed pair of values $(\dtf,\dtd)$, the flatband filling factor $\dtf$ defines statistical properties of the effective potential $\heps_{m,\beta}$ in~\eqref{eq:Ham_D_eff}, and consequently the behavior and the properties of the mobile interacting particles in the dispersive component. The interaction $V$ and the two filling fractions $\dtf,\dtd$ are the three control parameters which can drastically change the transport properties of the considered system. In particular, varying $V$ and $\dtd$ can lead to strong correlations, while varying $V$ and $\dtf$ will control the strength of effective disorder. We therefore expect MBL-like properties, despite the fact that the overall system is translationally invariant. Our considerations apply to systems with any number of single particle bands $\nu$, in any spatial dimension, and for any type of many-body statistics.

\begin{figure}
    \centering
    \includegraphics[width=\columnwidth]{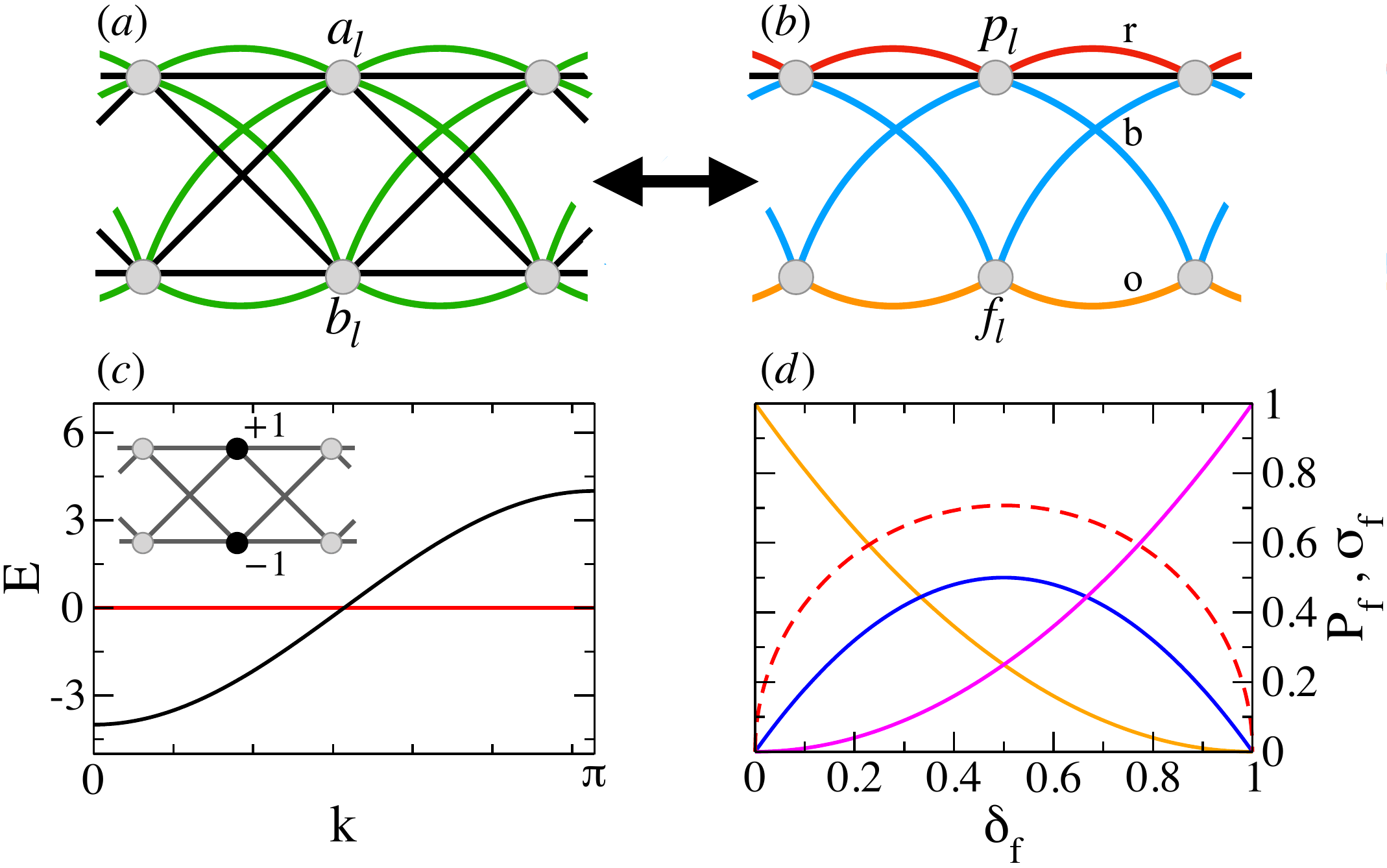}
    \caption{(a) The network~(\ref{eq:case1_sp},\ref{eq:case1_int}) with $\mhsp$ (straight black) and $\mhi$ (curved green). (b) Detangled $\mh$ network with $\mhsp$ (straight black), $\mhi^\text{d}$  [curved red (r)], $\mhi^\text{df}$ [curved blue (b)] and $\mhi^\text{f}$  [curved orange (o)]. (c) Band structure of $\mhsp$~\eqref{eq:case1_sp}, with the inset indicating the nonzero amplitude locations of an $E=0$ FB CLS (black filled circles). 
(d) Probabilities $P_\text{f}(\varepsilon_l=0)$ (orange), $P_\text{f}(\varepsilon_l=V)$ (blue), $P_\text{f}(\varepsilon_l=2V)$ (magenta) and  standard deviation $\sigma_\text{f}$ (red dashed) of $\heps_l$ in Eq.~\eqref{eq:case1_pot}  for $V=1$ versus $\dtf$.}
    \label{fig:ex_nu2}
\end{figure}

\sect{Signatures of many-body localization transition} 
As an example we consider spinless fermions in a one-dimensional network with two sites per unit cell, $\nu=2$. The Hamiltonian $\mh = \mhsp + V \mhi$ reads
\begin{align}
\small 
    \label{eq:case1_sp}
    \mhsp &= \sum_l - \big[ (\ha_l^\dagger + \hb_l^\dagger)(\ha_{l+1} + \hb_{l+1}  )+ \text{h.c.} \big], \\
    \mhi  &= \sum_l  \bigl[ \hn_{a,l} +  \hn_{b,l} \bigr] \bigl[ \hn_{a,l+1} +  \hn_{b,l+1}\bigr] .
    \label{eq:case1_int}
\end{align}
The actions of $\mhsp$ and $\mhi$ are shown in Fig.~\ref{fig:ex_nu2}(a) in black straight lines and green curves respectively. Both single particle hoppings and interactions connect all sites in neighboring unit cells. Nevertheless the single particle spectrum of $\mhsp$ consists of one dispersive band $E({\bf k}) = -4 \cos({\bf k})$ and one flatband $E=0$ with its orthonormal CLS shown in Fig.~\ref{fig:ex_nu2}(c). 

The local unitary transformation $\ha_l = (\hp_l + \hf_l)/\sqrt{2} $ and $\hb_l = (\hp_l - \hf_l)/\sqrt{2}$ introduced in Ref.~\onlinecite{flach2014detangling} detangles the Hamiltonian $\mhsp$~\eqref{eq:case1_sp} in $\mhsp^\text{d} = -2\sum_l \bigl[\hp_l^\dagger\hp_{l+1} + \text{h.c.}\bigr]$ with $\mhsp^\text{f}=0$. Moreover, the interaction $\mhi$~\eqref{eq:case1_int} is invariant under that rotation. In this new basis $\mhi$ breaks down into three components $\mhi^\text{f} =  \sum_l  \hn_{f,l} \hn_{f,l+1}$, $\mhi^\text{d} =  \sum_l  \hn_{p,l} \hn_{p,l+1}$, and $\mhi^\text{df} = \sum_l \bigl[ \hn_{p,l} \hn_{f,l+1} + \hn_{f,l} \hn_{p,l+1}\bigr]$ which are shown in Fig.~\ref{fig:ex_nu2}(b) with orange, red and blue curves respectively.

The Hamiltonian $\mhq$~\eqref{eq:Ham_D_eff} and its potential $\heps_l $ read
\begin{align}
    \label{eq:case1_effH}
    \hspace*{-2.5mm}  
    \mhq & = \sum_l \big[\heps_l \hn_{p,l} - 2(\hp_l^\dagger\hp_{l+1} + \text{h.c.}) + V \hn_{p,l} \hn_{p,l+1} \big] \\
    &\qquad \heps_l = V (\hq_{l+1} + \hq_{l-1}),
    \label{eq:case1_pot}
\end{align} 
with $\hn_{p,l}=\hp_l^\dagger\hp_l$. The conserved quantities $\hq_{l} = \hf_l^\dagger\hf_{l}$ take the values $\{ 0,1\}$. Hence $\mhq$ describes interacting spinless fermions in a one-dimensional chain with a random ternary potential $\varepsilon_l\in \{0, V, 2V\}$. Ternary disorder with equal probabilities $1/3$ has been studied in Ref.~\onlinecite{janarek2018discrete} where an MBL transition was reported for the Heisenberg spin-$1/2$ chain by varying the strength of interaction and disorder independently. Our case is trickier, since both disorder and interaction strengths are tuned by the same control parameter $V$ in Eq.~\eqref{eq:case1_effH}. Further, the probabilities of $\varepsilon_l =  \{0, V, 2V\}$ depend on the filling fraction $\dtf$ of the flatband component: $P_\text{f}(\varepsilon_l=0) = (1 - \dtf)^2$, $P_\text{f}(\varepsilon_l=V) = 2 \dtf (1 - \dtf)$ and $P_\text{f}(\varepsilon_l=2V)=  \dtf^2$. The curves are shown in Fig.~\ref{fig:ex_nu2}(d) including the standard deviation $\sigma_\text{f} = V \sqrt{2 \dtf (1 - \dtf)}$. The average potential value $\langle \varepsilon_l \rangle_l = 2V \dtf$. Note as well that equal probabilities $1/3$ are never realized for any filling fraction value.

We identify the transition between ergodic (thermalized, metallic, delocalized) and non-ergodic (non-thermalized, insulating, localized) regimes of our system by computing the energy-resolved adjacent gap ratio $r^{(n)} = \min(s^{(n)}, s^{(n+1)})/\max(s^{(n)}, s^{(n+1)})$ with $s^{(n)} = E_n - E_{n-1}$ for the eigenenergies $E_n$~\cite{oganesyan2007localization}. The expectation is that the ergodic regime corresponds to the Gaussian Orthogonal Ensemble (GOE) with $r_\text{GOE} = 0.5307$~\cite{atas2013distribution}. At variance, the non-ergodic regime should yield a Poisson distribution of level spacings with $r_\text{Poisson} \approx 0.3863$.

\begin{figure}
    \centering
    \includegraphics[width=\columnwidth]{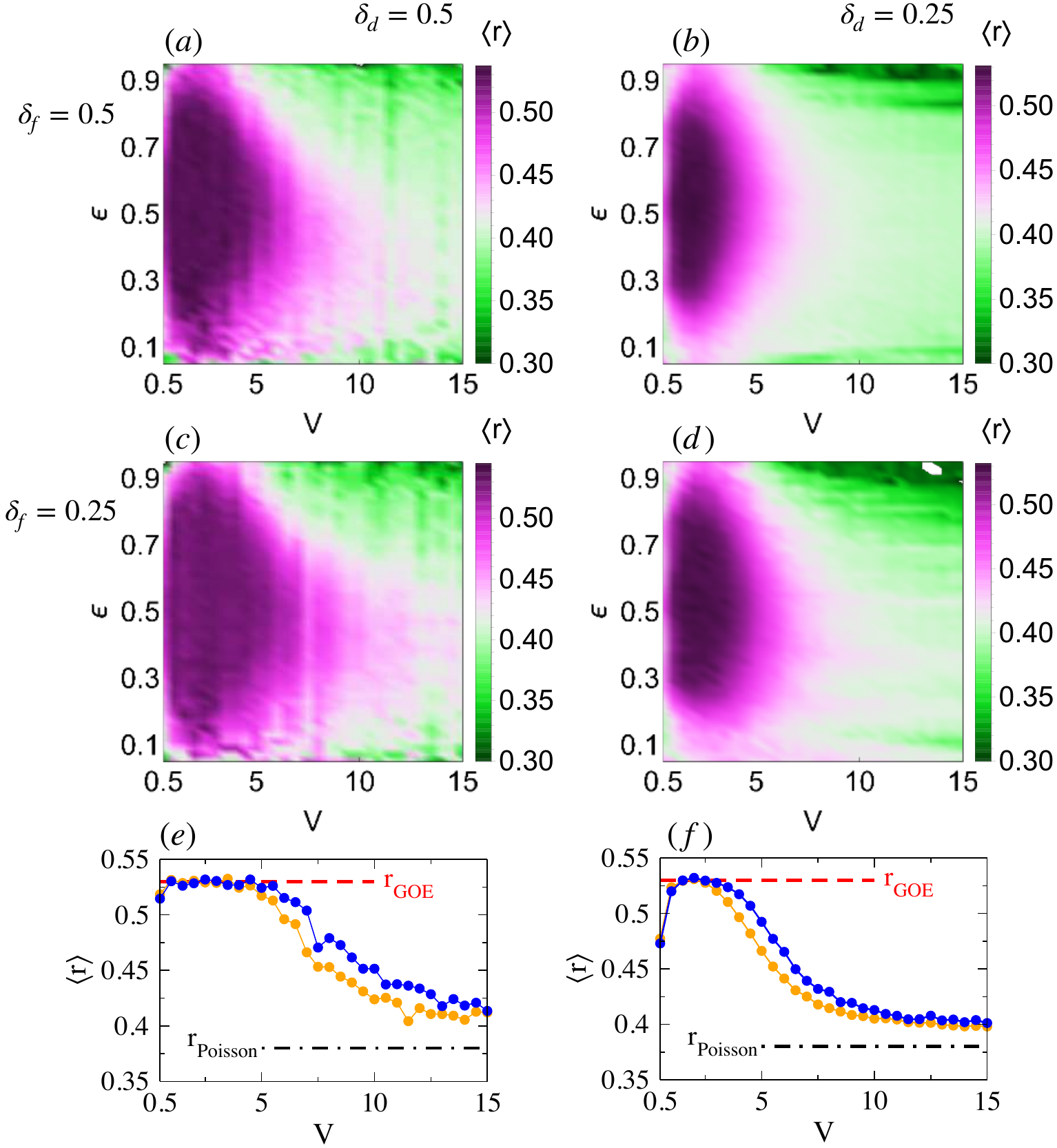}
    \caption{(a-d)  Energy resolved mean adjacent gap ratio $\langle r\rangle$ versus $(\epsilon,V)$ for model~(\ref{eq:case1_effH},\ref{eq:case1_pot}). (a) $\dtd=0.5;\dtf=0.5$; (b) $\dtd=0.25;\dtf=0.5$; (c) $\dtd=0.5;\dtf=0.25$; (d) $\dtd=0.25;\dtf=0.25$. (e) $\langle r\rangle$ versus $V$ for $\epsilon= 0.5$ at $\dtf=0.5$ (orange) and $\dtf=0.25$ (blue) for $\dtd=0.5$. (f) Same as (e) for $\dtd=0.25$. The system size $L=16$ for all cases.}
    \label{fig:ME_ex_nu2}
\end{figure}

We diagonalize $\mhq$ in Eq.~\eqref{eq:case1_effH} for $L=16$ sites with open boundary conditions averaging over $200$ realizations at fixed filling fractions $\dtf$ and $\dtd$ --  i.e. over 200 sectors of the Hilbert space~\footnote{No changes in the mobility edge profiles appeared while averaging over a different number of realizations}. Following Ref.~\cite{luitz2015many}, the spectrum is normalized as $\epsilon(E_n) = (E_n-E_\text{min})/(E_\text{max} - E_\text{min})$ for each realization, divided into $50$ intervals, and the mean adjacent gap ratio $\langle r\rangle$ is computed for each segment separately. The results are reported in Fig.~\ref{fig:ME_ex_nu2} (a-d). In all cases, the MBL transition emerges at large enough interaction strength $V\gg1$.
Note that the MBL transition occurs for different values of $V$ within different irreducible sectors of the Hilbert space characterized by different pairs of filling fractions $(\dtf,\dtd)$ despite sharing the same global filling $\delta$ -- {\t e.g.} Figs.~\ref{fig:ME_ex_nu2} (b,c). 

\begin{figure}
    \centering
    \includegraphics[width=\columnwidth]{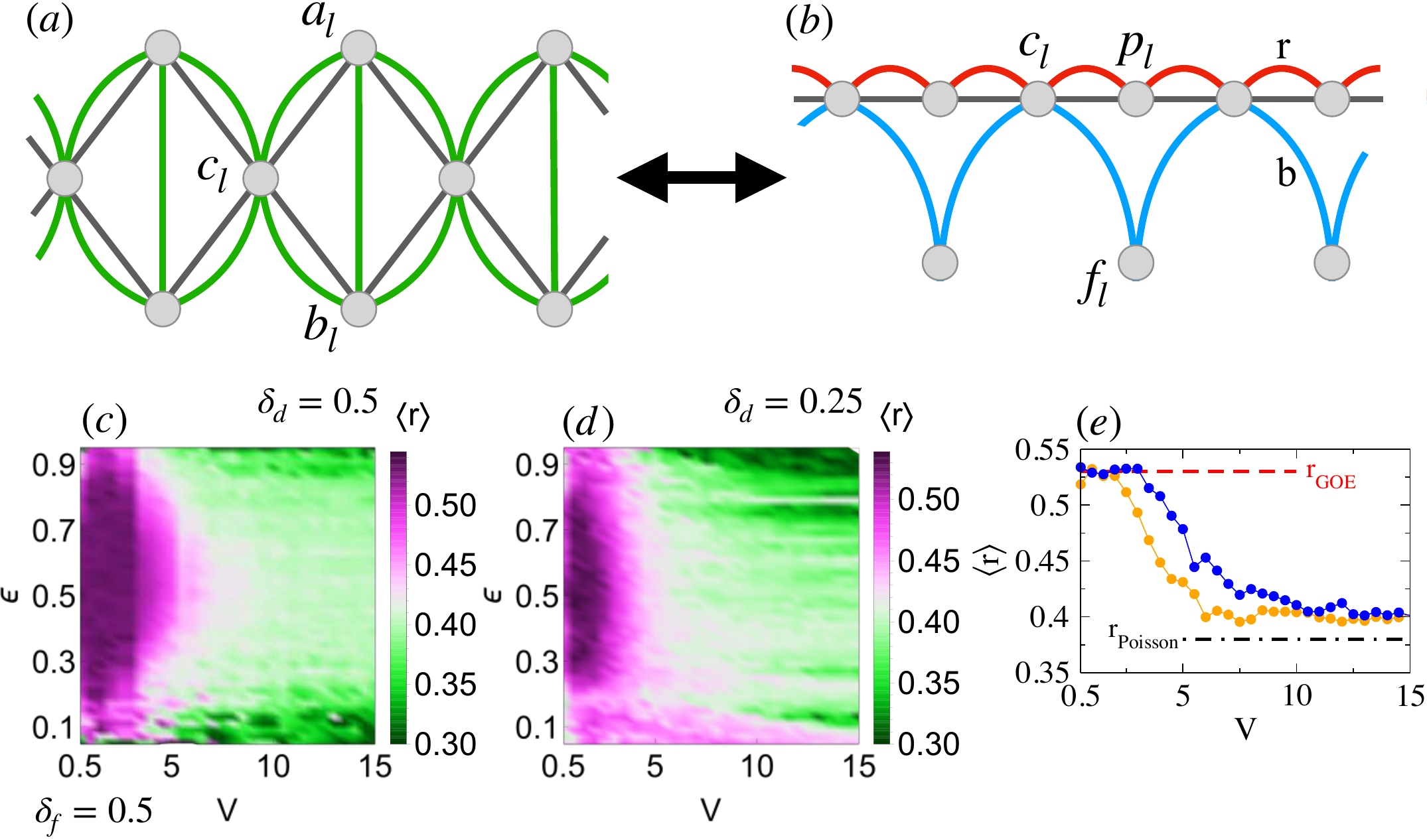}
    \caption{(a) The network~(\ref{eq:case2_sp},\ref{eq:case2_int}) with $\mhsp$ (straight black) and $\mhi$ (curved green). (b) Detangled network ~\eqref{eq:case2_effH} with  $\mhi^\text{d}$ [red (r)] and $\mhi^\text{df}$ [blue (b)]. (c-d) Energy resolved mean adjacent gap ratio $\langle r\rangle$ versus $(\epsilon,V)$ for (c) $\dtd=0.5;\dtf=0.5$;  (d) $\dtd=0.25;\dtf=0.5$. (e) $\langle r\rangle$ versus $V$ around $\epsilon= 0.5$ at $\dtd=0.5$ (blue) and $\dtd=0.25$ (orange) for $\dtf=0.25$. The system size $L=16$ for all cases.}
    \label{fig:ex_nu3}
\end{figure}

Our construction also explains the non-ergodic dynamics observed for a $\nu=3$ case with spinless fermions in Ref.~\onlinecite{daumann2020manybody}. The model is shown in Fig.~\ref{fig:ex_nu3}(a) and it is described by the Hamiltonian $\mh = \mhsp + V \mhi$ with
\begin{align}
    \small 
    \label{eq:case2_sp}
    \mhsp &= -\sum_l \big[(\ha_l^\dagger + \hb_l^\dagger)(\hc_{l} + \hc_{l+1}) + \text{h.c.} \big]\\
    \mhi &= \sum_l \bigl[\hn_{a,l} + \hn_{b,l}\bigr]\bigl[\hn_{c,l} + \hn_{c,l+1}\bigr] + \hn_{a,l} \hn_{b,l}
    \label{eq:case2_int}
\end{align}
The single particle spectrum of $\mhsp$ consists of two dispersive bands $E_{1,2}({\bf k}) = \pm 2\sqrt{2} \cos({\bf k}/2)$ and a flatband $E=0$ with orthonormal CLS. Detangling local unitary transformations for the single particle Hamiltonian have been reported in Ref.~\onlinecite{flach2014detangling}. For spinless fermions, the resulting system is shown in Fig.~\ref{fig:ex_nu3}(b). In the new basis the product terms in Eq.~\eqref{eq:case2_int} decompose into $\mhi^\text{f} =  0$, $\mhi^\text{d} =  \sum_l   \hn_{p,l} (\hn_{c,l} +  \hn_{c,l+1})$, and $\mhi^\text{df} = \sum_l \hn_{f,l} (\hn_{c,l} + \hn_{c,l+1})$ -- defining the Hamiltonian
\begin{align}
    \label{eq:case2_effH}
    \mhq &= \sum_l \big[ \heps_l \hn_{c,l} - \sqrt{2} (\hp_l^\dagger\hc_l + \hp_l^\dagger\hc_{l+1} + \text{h.c}) \\
    &\qquad + V \hn_{p,l} (\hn_{c,l} + \hn_{c,l+1})\big]\notag
\end{align}
with the potential $\heps_l = V (\hq_{l-1} + \hq_{l} )$. Interestingly, the terms $\hn_{a,l} \hn_{b,l}$ result in $\hp_l^\dagger\hp_l^\dagger\hf_l\hf_l +\text{h.c}$ in addition to the Hubbard interaction terms on $p,f$ sites which could violate the finetuning. These terms need more than one particle per state and therefore disappear for spinless fermions. Consequently they do not enter $\mhq$ in Eq.~\eqref{eq:case2_effH}. However such terms will appear for \emph{e.g.} bosons and spinful fermions, and will move pairs of particles between sublattices $\cF$ to $\cD$. Consequently the quantities $\hq_{l}$ will be no longer conserved for spinful fermions or bosons, and will wash out the irreducible sectors in the Hilbert space of $\mh$. In Fig.~\ref{fig:ex_nu3} (c-d) we plot the energy-resolved mean adjacent gap ratio $\langle r\rangle$ for two pairs of filling factors $\dtd$ and $\dtf$ upon increasing the interaction strength $V$. We observe signatures of an MBL transition for large $V\gg1$. The transition is further visualized in Fig.~\ref{fig:ex_nu3} (e) where we plot the energy-resolved mean $\langle r\rangle$ versus $V$ around $\epsilon= 0.5$.

\sect{Conclusions}
To conclude, we showed that disorder free many body localization is obtained for flatband networks with finetuned interaction. The flatband supports compact localized states, and the finetuning locks particles in these states even in the presence of interaction. These locked particles turn into scatterers for particles from dispersive states. These families have been obtained by fine-tuning two--body interaction terms on single particle lattices that host dispersive bands and flat bands with orthonormal sets of CLS. We showed that these scatterers are equivalent to conserved quantities and  enter the Hamiltonian of the system inducing an effective disorder. We studied numerically two sample cases, Eqs.~(\ref{eq:case1_sp},\ref{eq:case1_int}) and Eqs.~(\ref{eq:case2_sp},\ref{eq:case2_int}), for spinless fermions in 1D, confirming  that such disorder indeed induces many-body localization transition upon changing the interaction strength. The proposed fine-tuning scheme applies in any lattice dimensions and for any type of many-body statistics. We therefore arrive at a systematic generic generator of quantum many-body systems characterized by an extensive number of local conserved operators, which result in ergodicity breaking phenomena. Another important extension is that we can abandon the translational invariance of $\mh$ and consider local rotations and/or the energies $E_a$ in Eq.~\eqref{eq:Ham_FB} unit cell dependent; one can also consider flatband Hamiltonians $\mhsp$ without orthonormal sets of CLS~\cite{leykam2017localization}.

Our findings explain recent spinless fermion results for a rhombic lattice considered by Daumann \emph{et.al.}~\cite{daumann2020manybody,khare2020localized} (orthonormal set of CLS) as well as for a sawtooth ladder considered by Khare {\it et.al.}~\cite{khare2020localized} (non-orthonormal set of CLS). It is straightforward to observe that for total filling fraction $\delta\leq 1/\nu$ there are exact eigenstates with all particles confined to single particle CLS. These eigenstates coexist with extended eigenstates characterized by volume-law entanglement, and become flatband many-body quantum scars~\cite{hart2020compact,mcclarty2020disorder,kuno2020flat_qs,danieli2020quantum}. Many-body quantum scars are related to weak ergodicity breaking phenomena~\cite{turner2018weak,serbyn2020quantum,Pilatowsky2021ubiquitous}. 

The above obtained fine-tuned models can be both appealing from a purely mathematical point of view, and for experimentally relevant setups. Indeed, flatband networks have been indeed experimentally realized in diverse platforms, such as ultra cold atoms~\cite{taie2015coherent} and photonic lattices~\cite{mukherjee2015observation,vicencio2015observation,weimann2016transport} -- see  also~\cite{derzhko2015strongly,leykam2018artificial,leykam2018perspective}. Flatband systems with orthonormal CLS allow the energy levels $\efb$ in Eq.~\eqref{eq:Ham_FB} to be freely tuned. They can thus either cross the dispersive bands $E_a(\mk)$ in Eq.~\eqref{eq:Ham_DB} or be gapped away from dispersive bands. This freedom allows to tune the flatband energy at the Fermi level, and to load the particles into the flatband states prior reaching the complete filling of dispersive bands in an experimentally achievable way. 

\sect{Acknowledgments}
This work was supported by the Institute for Basic Science (Project number IBS-R024-D1). We thank I. Khaymovich for helpful discussions. 

\bibliography{general,flatband,mbl}

\end{document}